\documentstyle[spie]{article}

\def\degr{\hbox{$^\circ$}}
\def\fm{\hbox{$.\!\!^{\rm m}$}}
\def\farcs{\hbox{$.\!\!^{\prime\prime}$}}

\input{psfig} 

\title{Computer simulations of interferometric imaging with the VLT
       interferometer and the AMBER instrument} 

\author{T. Bl\"ocker, K.-H. Hofmann, F. Przygodda and G. Weigelt
\skiplinehalf 
Max-Planck-Institut f\"ur Radioastronomie, Auf dem H\"ugel 69, 53340 Bonn, Germany
}

\begin{document} 
  \maketitle 

%%%%%%%%%%%%%%%%%%%%%%%%%%%%%%%%%%%%%%%%%%%%%%%%%%%%%%%%%%%%% 
\begin{abstract}
We present computer simulations of interferometric imaging with the VLT
interferometer and the AMBER instrument. These simulations include both the
astrophysical modelling of a stellar object by radiative transfer
calculations and the simulation of  light propagation from the object to the
detector (through  atmosphere, telescopes, and the AMBER instrument),
simulation of photon noise and detector read-out noise, and finally data
processing of the interferograms. The results show the dependence of the
visibility error bars on the following observational parameters:
different seeing during the observation of
object and reference star (Fried parameters $r_{0,{\rm object}}$=2.4\,m,
$r_{0,{\rm ref.}}$=2.5\,m), different residual tip-tilt error
($\delta_{\rm tt,object}$=2\% of the Airy disk diameter,
$\delta_{\rm tt,ref.}$=0.1\%), and object 
brightness ($K_{\rm object}$=3.5\,mag and 11\,mag, $K_{\rm ref.}$=3.5\,mag).
Exemplarily, we
focus on stars in late stages of stellar evolution and study one of its
key objects, the dusty supergiant IRC\,+10\,420 that is rapidly evolving on
human timescales.
We show computer simulations of VLTI interferometry of  IRC\,+10\,420 with
two ATs (wide-field mode, i.e.\ without fiber optics spatial filters)
and discuss whether the visibility accuracy is
sufficient to distinguish between different theoretical model predictions.
\end{abstract}

%>>>> Please include a list of keywords after the abstract 

\keywords{VLTI, AMBER, interferometric imaging, IRC\,+10\,420, dust shells, 
          circumstellar matter, supergiants, stellar evolution}

%%%%%%%%%%%%%%%%%%%%%%%%%%%%%%%%%%%%%%%%%%%%%%%%%%%%%%%%%%%%%
\section{INTRODUCTION}
\label{sect:intro}  
The Very Large Telescope Interferometer\cite{GlinEtal2000}
(VLTI) with its four 8.2\,m unit
telescopes (UTs) and three 1.8\,m auxiliary telecopes (ATs) will certainly
establish
a new era of optical and infrared interferometric imaging within the next few
years. With a maximum baseline of up to more than 200\,m,
the VLTI will allow the study of astrophysical key objects with
unprecendented resolution opening up new vistas to a better understanding of
their physics. 

The near-infrared focal plane instrument of the VLTI,
the Astronomical MultiBEam Recombiner\cite{PetEtal98,PetEtal2000}
(AMBER),
will operate between 1 and 2.5\,$\mu$m and for up to three beams allowing
the measurement of closure phases. In a second phase
its wavelength coverage is planned to be extended to 0.6\,$\mu$m.
Objects as faint as  $K=20$\,mag are expected to be observable with
AMBER when a bright reference star is available, and as faint as
$K=14$\,mag otherwise.

Among the astrophysical key issues\cite{RichEtal2000}
are, for instance,
young stellar objects, active galactic nuclei and stars in late
phases of stellar evolution. %In our simulations, we will focus on the latter.
The simulation of a stellar object
consists in principle of two components: \\[0.5ex]
%%%\begin{itemize}
%%%\item[(i)]
\begin{minipage}[t]{5ex}
\hspace*{1ex} (i)
\end{minipage}
\begin{minipage}[t]{0.95\textwidth}
the calculation of an astrophysical model of the object, typically based on
           radiative transfer calculations predicting,
           e.g., its intensity distribution.
           To obtain a robust and non-ambiguous model,
           it is of particular importance to take 
           {\it diverse} observational constraints into account, for instance 
           the spectral energy distribution and visibilities.
\end{minipage} \\[0.3ex]
%\item[(ii)]
\begin{minipage}[t]{5ex}
\hspace*{1ex} (ii)
\end{minipage}
\begin{minipage}[t]{0.95\textwidth}
the determination of the interferometer's response to this intensity signal,
i.e.\ the simulation of light propagation in the atmosphere and the
interferometer.
%%%            and the simulation of Michelson interferograms of the object.
\end{minipage} \\[1.0ex]
%%            disturbances by the earth's atmosphere.
%\end{itemize}
Often, only one of the above parts is considered in full detail.
%%%i.e. {\it either} to match the observations by astrophysical models
%%%{\it or} to calculate the impact of the atmosphere and the telescope optics
%%%on the appearence of a given intensity distribution.
The aim of this study is to combine both efforts and to present a computer
simulation
of the VLTI performance for observations of one object class, the dusty
supergiants. For this purpose, we calculated a detailed radiative transfer
model for one of its most outstanding representatives, the supergiant
IRC\,+10\,420, and carried out
computer simulations of VLTI visibility measurements.
The goal is to estimate how
accurate visibilities can be measured with the VLTI in this particular
but not untypical case, to discuss if the accuracy is sufficient to distinguish
between different theoretical model predictions and to study on what the
accuracy is dependent.

\section{The supergiant IRC\,+10\,420: Evolution on human timescales}
The star IRC\,+10\,420
(= V\,1302~Aql = IRAS\,19244+1115) is an
outstanding object for the study of stellar evolution.
Its spectral type changed from
F8\,I$_{\rm a}^{+}$ in 1973\cite{HumStrMurLow73} 
to mid-A today\cite{OudGroeMatBloSah96,KloChePan97}
corresponding to an effective temperature increase
of 1000-2000\,K within only 25\,yr.
It is one of the
brightest IRAS objects
due to its very strong infrared excess by circumstellar dust 
and one of the warmest stellar OH maser sources
known\cite{GigWooWeb76,MutEtal79,NedBow92}.
Large mass-loss rates, typically of the
order of several $10^{-4}$\,M$_{\odot}$/yr
were determined by CO\,observations\cite{KnaMor85,OudGroeMatBloSah96}.
IRC\,+10\,420 is believed to be a massive luminous star currently being
observed in its rapid transition from the red supergiant stage to the
Wolf-Rayet phase\cite{MutEtal79,NedBow92,JonHumGehEtal93,OudGroeMatBloSah96,KloChePan97}.
Its massive nature (initially $\sim 20$ to 40 M$_{\odot}$)
can be concluded from its
distance ($d$ = 3--5 kpc) and large wind velocity (40 km/s)
ruling out alternative scenarios. IRC\,+10\,420 is the {\it only}
object observed until now in its transition to
the Wolf-Rayet phase.
The structure of the circumstellar environment of
IRC\,+10\,420 appears to be very complex\cite{HumSmiDavEtal97},
and scenarios proposed
to explain the observed spectral features of IRC\,+10\,420
include a rotating equatorial disk\cite{JonHumGehEtal93}, 
bipolar outflows\cite{OudEtal94}, and
the infall of circumstellar material onto the star's photosphere
\cite{Oud98}. 

Several infrared speckle and coronogra\-phic
observations
\cite{DyckEtal84,RidgEtal86,CobFix87,ChrEtal90,KastWein95}
were conducted to study the dust shell of IRC\,+10\,420.
The most recent study\cite{BloeckEtal99}
reports the first diffraction-limited  73\,mas
bispectrum speckle interferometry of IRC\,+10\,420
and presents the first radiative transfer calculations
that model {\it both} the spectral energy distribution
{\it and} the visibility of this key object.
In the following we will briefly describe
the main results and conclusions of this study which will serve as
astrophysical input for the VLTI computer simulations presented in the
next section. 

\subsection{Visibility measurements and bispectrum speckle interferometry}

Speckle interferograms of IRC\,+10\,420 
were obtained with the Russian
6\,m telescope at the Special Astrophysical Observatory
on June 13 and 14, 1998\cite{BloeckEtal99}.
The speckle data were recorded
with our NICMOS-3 speckle camera
through an interference filter with a
centre wavelength of 2.11\,$\mu$m and a bandwidth of 0.19\,$\mu$m.
The observational parameters were as follows:
exposure time/frame 50~ms; number of frames 8400;
2.11\,$\mu$m seeing (FWHM) $\sim$1\farcs0.
A diffraction-limited image of IRC\,+10\,420
with 73\,mas resolution was reconstructed from
the speckle interferograms using the bispectrum speckle interferometry
method\cite{Wei77,LohWeiWir83,HofWei86}.
The modulus of the object Fourier transform (visibility) was
determined  with the speckle interferometry method\cite{Lab70}.
\begin{figure}[hptb]
\begin{center}
  \begin{tabular}{c}
    \psfig{figure=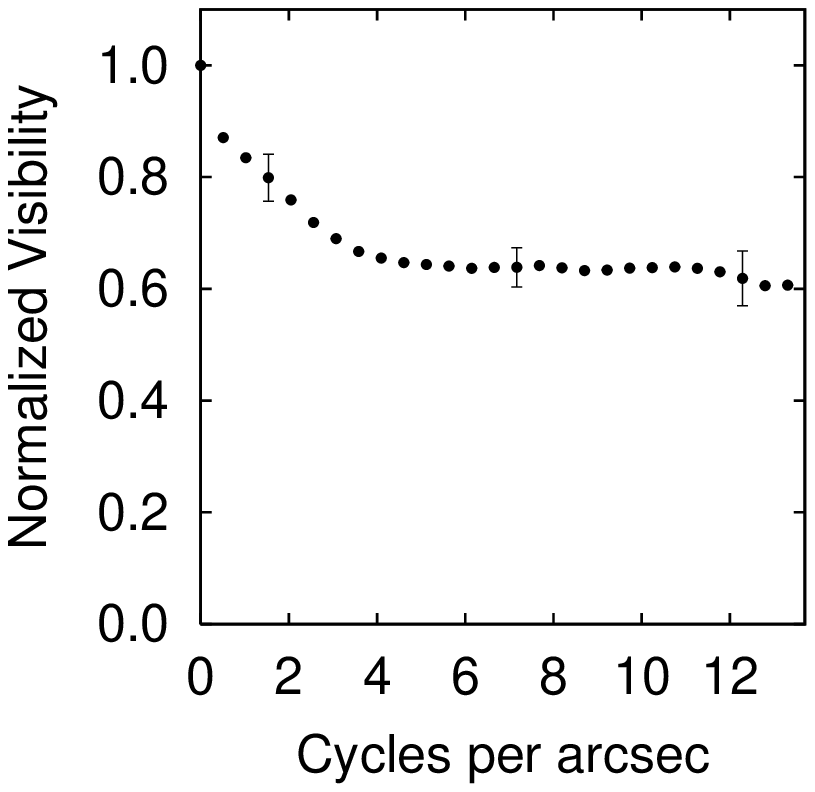,height=6cm} 
    \psfig{figure=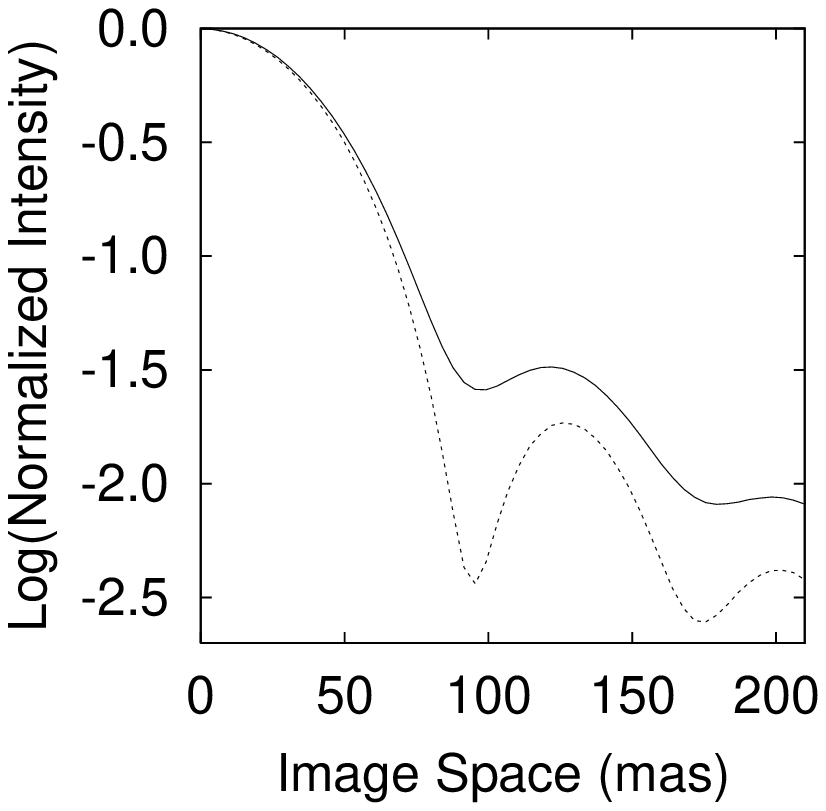,height=6cm} 
  \end{tabular}
\end{center}
\caption{
{\bf Left:}
Azimuthally averaged  2.11\,$\mu$m visibility of IRC\,+10\,420 
with error bars for selected frequencies.
This visibility function consists of a constant plateau (visibility $\sim 0.6$)
caused by the unresolved central object
and a triangle-shaped low-frequency
function caused by the faint extended nebula.
{\bf Right:}
Azimuthally averaged radial plots of the reconstructed diffraction-limited
$2.11 \mu$m-images of
IRC\,+10\,420 (solid line) and HIP\,95447 (dashed line).
} \label{Fvisi}
\end{figure}
Figure~\ref{Fvisi} (left panel)
shows the reconstructed 2.11\,$\mu$m visibility function of
IRC\,+10\,420. There is only marginal evidence
for an elliptical visibility shape
(position angle of the long axis $\sim 130\degr \pm 20\degr$, 
axis ratio $\sim 1.0$ to 1.1).
The visibility 0.6 at frequencies $>4$\,cycles/arcsec
shows that the stellar contribution to the total flux is $\sim$\,60\%
and the dust shell contribution is $\sim$\,40\%.
The Gau\ss\ fit FWHM diameter of the dust shell was determined to be
$d_{\rm FWHM} = (219 \pm 30)$\,mas.
By comparison, previous 3.8\,m telescope K-band observations\cite{ChrEtal90}
found a dust-shell flux contribution of $\sim$50\% and
$d_{\rm FWHM} = 216$\,mas.
However, as will be shown later, 
a ring-like intensity distribution appears
to be much better suited than the assumption of a Gaussian distribution
whose corresponding FWHM diameter fit may give misleading sizes
(see Sect.~\ref{Sssdust}).
The right panel of Fig.~\ref{Fvisi} displays the azimuthally averaged
diffraction-limited images of IRC\,+10\,420 and the unresolved star HIP 95447.
The $K$-band visibility  will strongly constrain radiative 
transfer calculations as shown in the next section.
\subsection{Dust-shell models}  \label{Sssdust}
\begin{figure}[hptb]
\begin{center}
  \begin{tabular}{c}
    \psfig{figure=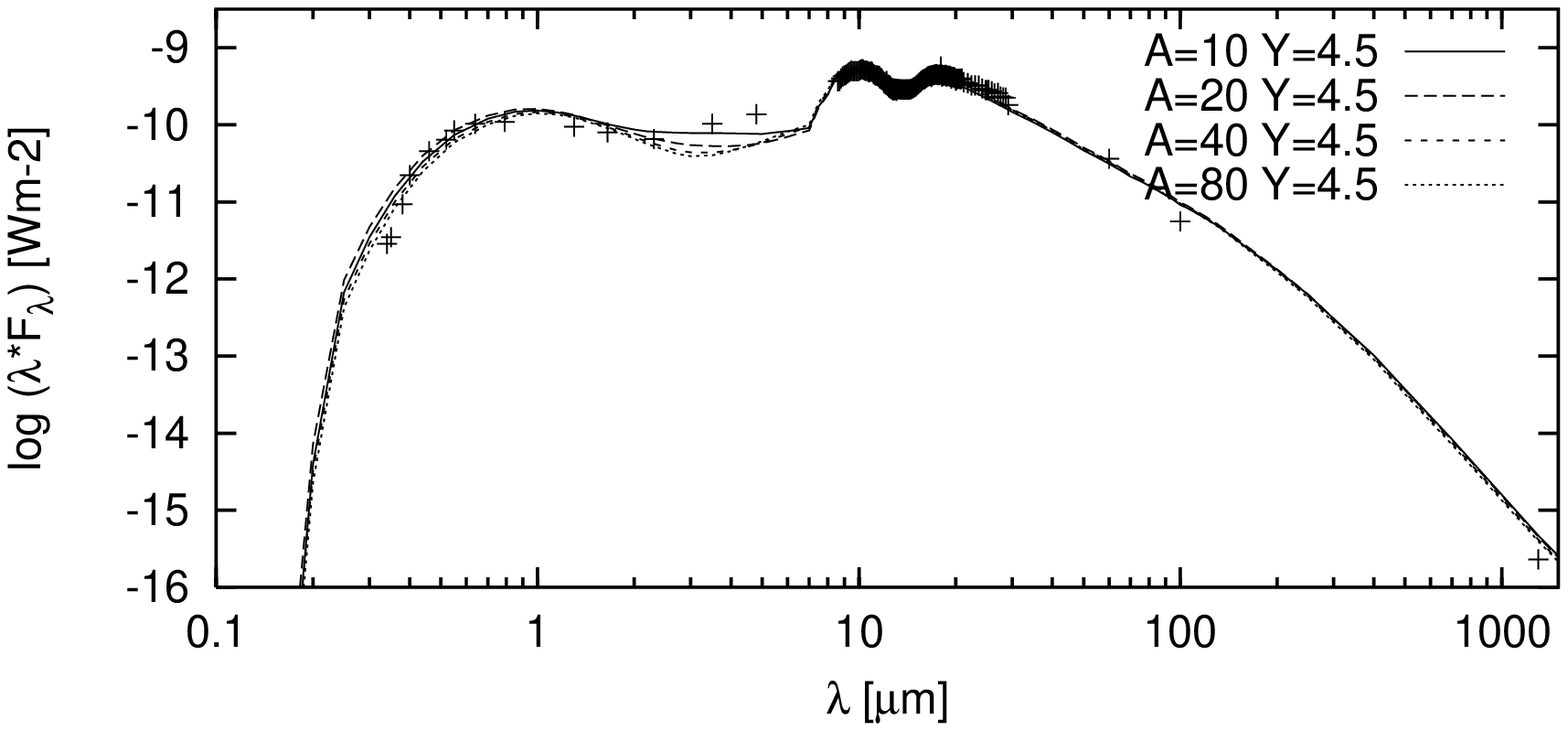,width=0.65\textwidth} \\
    \psfig{figure=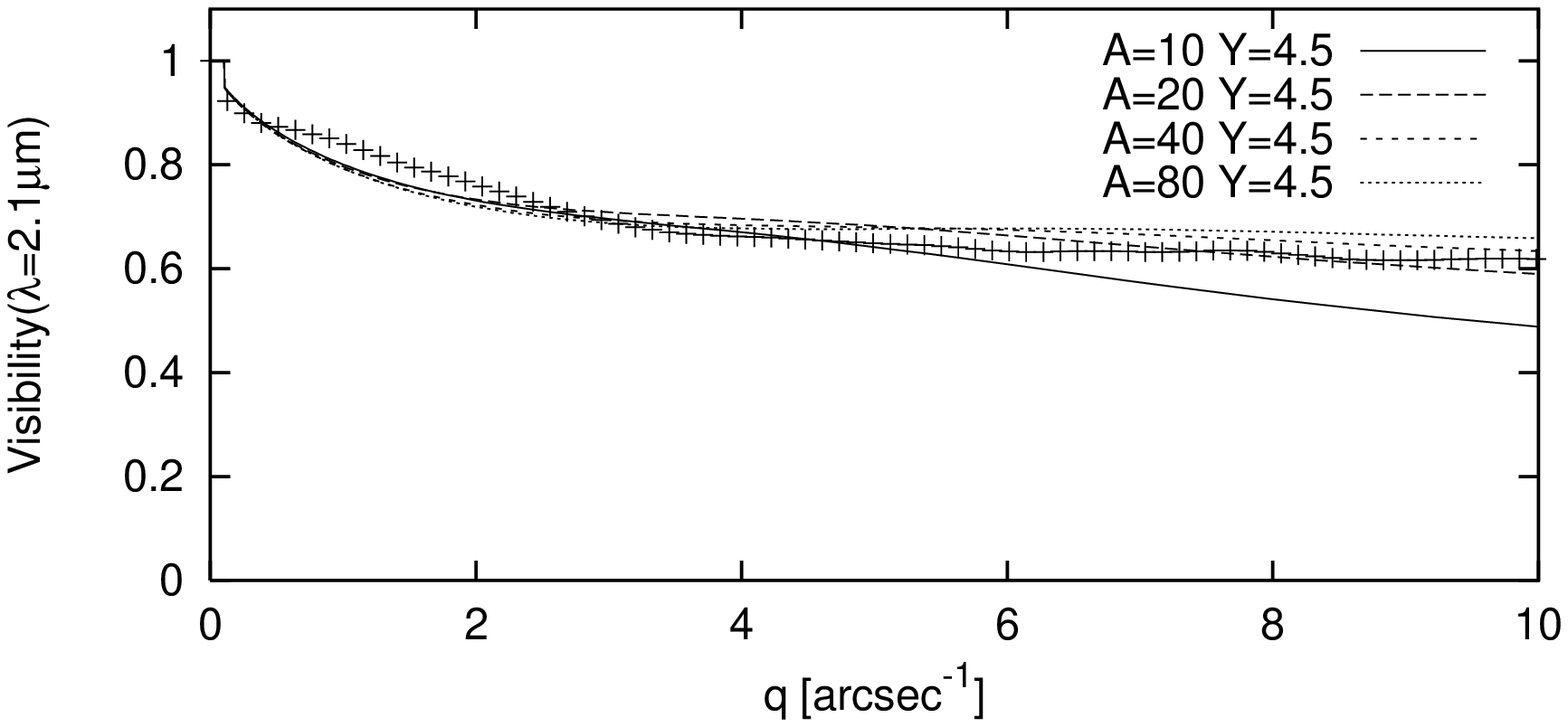,width=0.65\textwidth}  
\end{tabular}
\end{center}
\caption[Fsedvisidens17]
{
SED (top) and 2.11\,$\mu$m visibility (bottom)
for a superwind model with  $Y=r/r_{1}=4.5$ and different amplitudes. The
inner shell obeys a $1/r^{2}$ density distribution, the outer shell a
$1/r^{1.7}$ density distribution.
Model parameters are:
black body, $T_{\rm eff}=7000$\,K, $T_{1}=1000$\,K,
$\tau_{0.55\mu{\rm m}}=7.0$, silicates\cite{DraLee84},
standard grain size distribution\cite{MRN77} with
$a_{\rm max}=0.45\,\mu$m, and $Y_{\rm out}=10^{4}$.
The symbols
refer to the observations (see text) corrected for interstellar extinction of
$A_{\rm v}=5^{\rm m}$.
}                                      \label{Fsedvisidens17}
\end{figure}

The spectral energy distrubion (SED)
of IRC\,+10\,420 with
9.7 and 18\,$\mu$m silicate emission features
is shown in Fig.~\ref{Fsedvisidens17}.
It corresponds to the '1992' data set used
by Oudmaijer et al.\cite{OudGroeMatBloSah96})
and combines VRI (October 1991),
near-infrared (March and April 1992) and Kuiper Airborne Observatory
(June 1991) photometry\cite{JonHumGehEtal93} with
the IRAS measurements and 1.3\,mm data\cite{WalEtal91}.
Additionally, we included data
for $\lambda < 0.55\,\mu$m\cite{CraEtal76}.
In contrast to the near-infrared, the optical magnitudes have remained
constant during the
last twenty years within a tolerance of $\approx 0 \fm 1$.

IRC\,+10\,420 is highly reddened due to an extinction of
$A^{\rm total}_{\rm V} \approx 7^{\rm m}$
by the interstellar medium and the circumstellar shell.
From polarization studies an interstellar extinction of
$A_{\rm V} \approx 6^{\rm m}$ to $7^{\rm m}$ was estimated
\cite{CraEtal76,JonHumGehEtal93}.
Based on the strength of the diffuse interstellar bands, 
$E(B-V)=1 \fm 4  \pm  0 \fm 5$ can be inferred for the
interstellar contribution compared to a total of  $E(B-V)=2\fm 4$\cite{Oud98}.
We  will use an interstellar $A_{\rm V}$ of
$5^{\rm m}$
adopting the method of Savage \& Mathis~\cite{SavMat79} with
$A_{\rm V} = 3.1 E(B-V)$\cite{OudGroeMatBloSah96}.
%
%
%\begin{figure}[hptb]
%\begin{center}
%  \begin{tabular}{c}
%    \psfig{figure=h1438_tp.f10.ps,height=7cm} \\
%    \psfig{figure=h1438_bt.f10.ps,height=7cm}  
%\end{tabular}
%\end{center}
%\end{figure}

In order to model {\it both} the observed SED {\it and}
$2.11\,\mu$m visibility, 
we performed radiative transfer calculations for dust shells
assuming spherical symmetry.
We used the code DUSTY\cite{IveNenEli97} 
which solves the spherical radiative transfer problem utilizing the
self-similarity and scaling behaviour of IR emission from radiatively
heated dust\cite{IveEli97}.
To solve the radiative transfer problem including absorption, emission and
scattering several properties of the central source and its surrounding
envelope are required, viz. (i) the spectral shape of the central source's
radiation; (ii) the dust properties,
i.e. the envelope's chemical composition and grain size distribution
as well as the dust temperature at the inner boundary; (iii) the relative
thickness of the envelope, i.e. the ratio of outer to inner shell radius,
and the  density distribution; and (iv) the total optical depth at a
given reference wavelength.
The code has been expanded for the calculation of synthetic visibilities.

\begin{figure}%[hptb]
\begin{center}
  \begin{tabular}{c}
    \psfig{figure=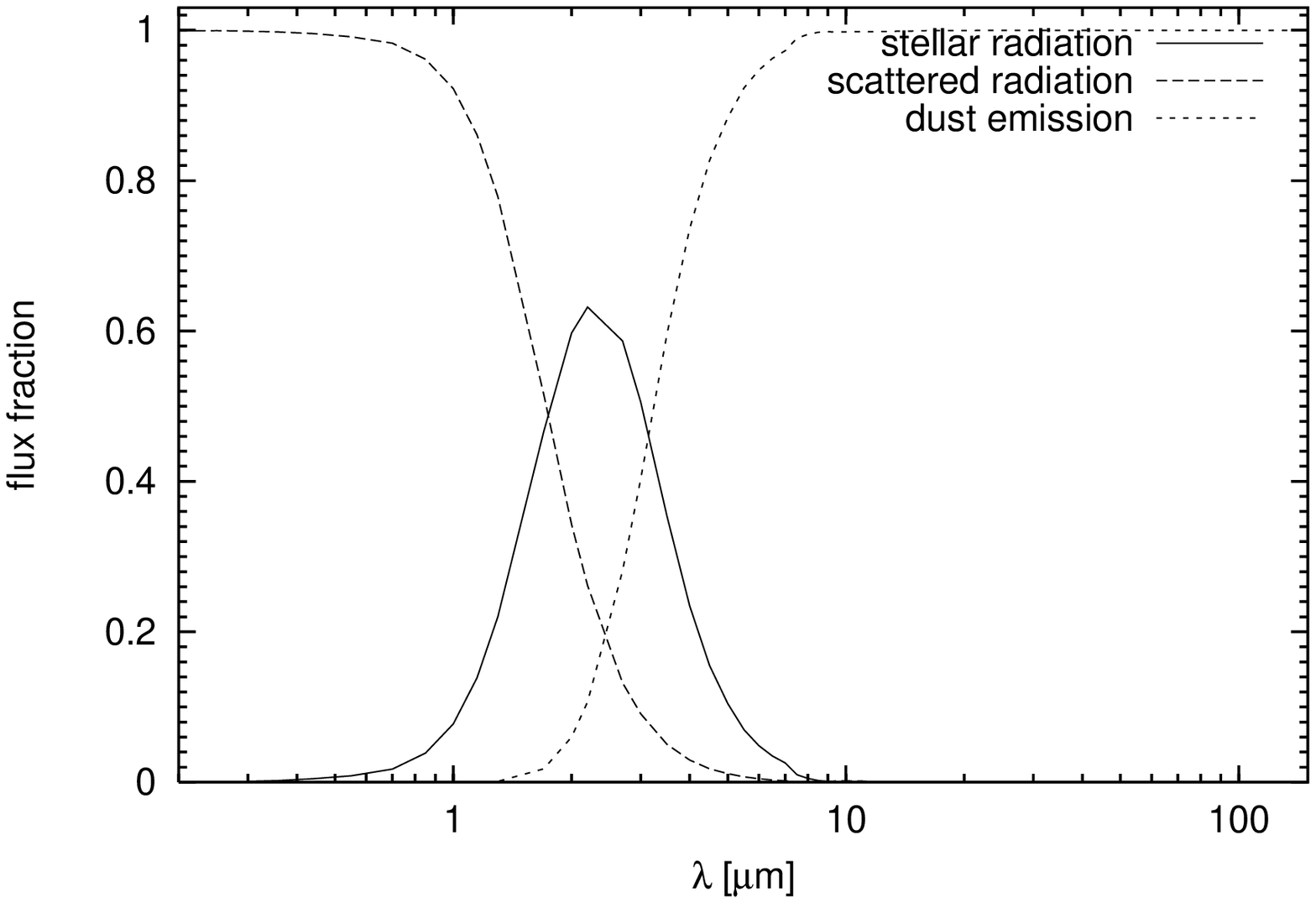,width=0.48\textwidth}
    \psfig{figure=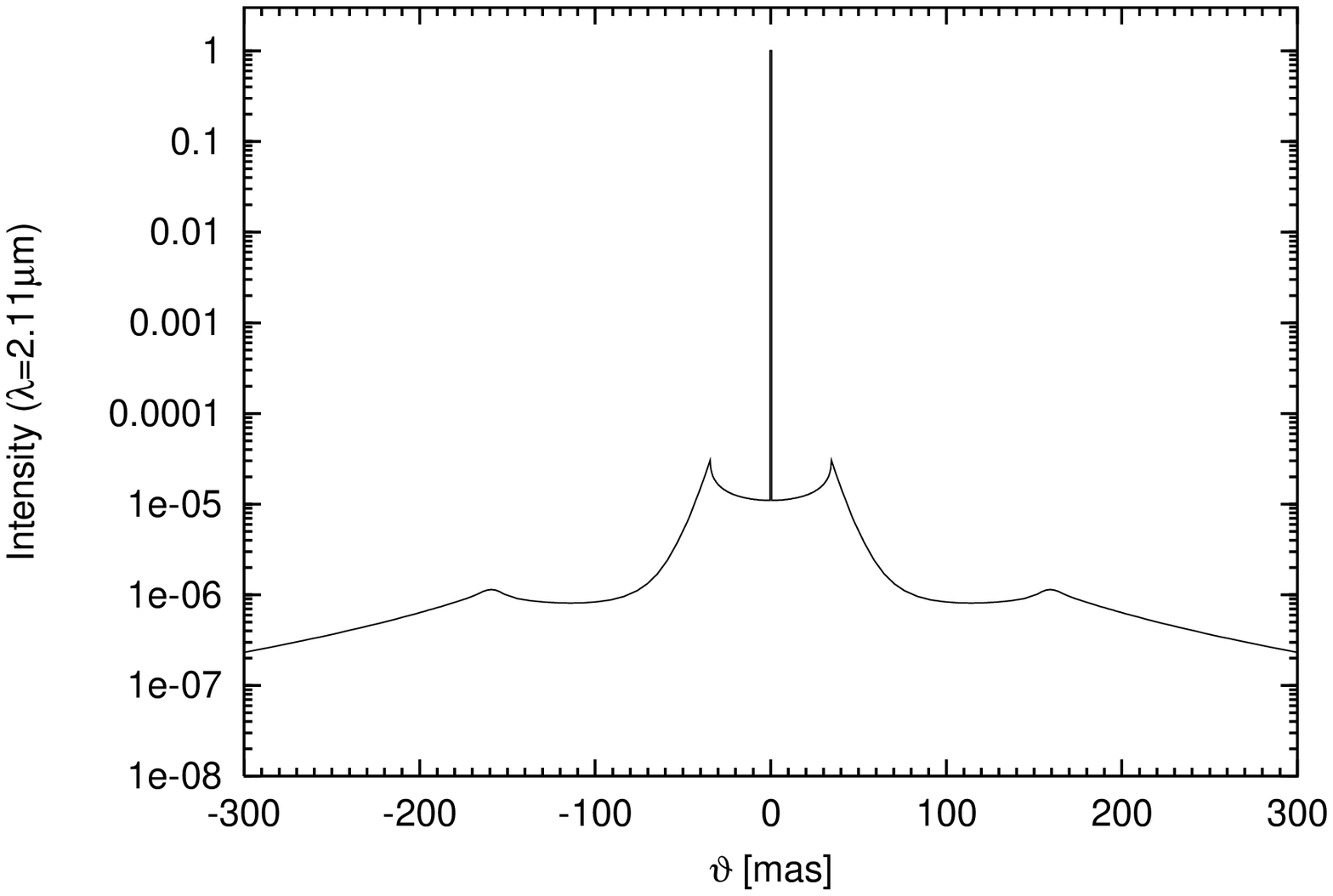,width=0.48\textwidth}  
\end{tabular}
\end{center}
\caption[ffraction17]
{
{\bf Left:}
 Fractional contributions of the emerging stellar radiation 
 as well as  of the scattered radiation and of the dust emission to the total
 flux as a function of the wavelength for a superwind model with
 $Y=r/r_{1}=4.5$, $A=40$ and a $1/r^{1.7}$ density distribution in the outer
 shell.
 Model parameters are:
 black body, $T_{\rm eff}=7000$\,K, $T_{1}=1000$\,K,
 $\tau_{0.55\mu{\rm m}}=7.0$, silicates\cite{DraLee84},
 standard grain size distribution\cite{MRN77} with
 $a_{\rm max}=0.45\,\mu$m, and $Y_{\rm out}=10^{4}$.
{\bf Right:}
 Corresponding
 normalized intensity vs.\ angular displacement $\vartheta$.
% for the superwind model with  $Y=r/r_{1}=4.5$, $A=40$ and 
% a $1/r^{1.7}$ density distribution in the outer shell.
 The (unresolved) central peak belongs to the central star.
 The inner hot rim of the circumstellar shell has a radius of 35\,mas, and
 the cool component is located at 155\,mas. Both loci correspond to local
 intensity maxima. 
% Model parameters are:
% black body, $T_{\rm eff}=7000$\,K, $T_{1}=1000$\,K,
% $\tau_{0.55\mu{\rm m}}=7.0$, Draine \& Lee (1984) silicates,
% Mathis et al.\ (1977) grain size distribution with $a_{\rm max}=0.45\,\mu$m.
}                                      \label{Fffraction17}
\end{figure}

We calculated various models of the dust shell of IRC\,+10\,420
considering black bodies and model atmospheres
as central sources of radiation, different silicates, grain-size and
density distributions. We refer to Bl\"ocker et al.~\cite{BloeckEtal99}
for a full description of the model grid. 
The remaining fit parameters are the  dust temperature, $T_{1}$, which
determines the radius of the shell's inner boundary, $r_{1}$,
and the optical depth, $\tau$, at a given reference wavelength,
$\lambda_{\rm ref}$. We refer to 
$\lambda_{\rm ref} =  0.55\,\mu$m. Models were calculated for
dust temperatures between 400 and 1000\,K and
optical depths between 1 and 12. Significantly larger
values for $\tau$ lead to silicate features in absorption.

The near-infrared visibility strongly constrains
dust shell models since it is, e.g., a sensitive indicator of the grain size.
Accordingly, high-resolution interferometry results provide essential
ingredients for models of circumstellar dust-shells.
In the instance of IRC\,+10\,420 (assuming the central star
to be a black body of $T_{\rm eff}=7000$\,K),
the silicate\cite{DraLee84} grain sizes, $a$, were found to be in
accordance with
a standard distribution function\cite{MRN77},
$n(a)$\,$\sim$\,$a^{-3.5}$, with $a$ ranging between
$a_{\rm min}$\,=\,0.005\,$\mu$m and $a_{\rm max}$\,=\,0.45\,$\mu$m.

However, the observed dust shell properties cannot be fitted by
single-shell models but seem to require multiple components.
At a certain distance we considered an enhancement over the assumed  $1/r{^x}$
density distribution.
The best model for {\it both} SED {\it and} visibility was found  
for a dust shell with a dust temperature of 1000\,K at its inner radius
of $r_{1}= 69\,R_{\ast}$. At a distance of
$r = 308\,R_{\ast}$ ($Y = r/r_{1} = 4.5$)
the density was enhanced by a factor of $A=40$ and and its
density exponent was changed from $x=2$ to $x=1.7$.
These fits for SED and  2.11\,$\mu$m visibility are shown in
Fig.~\ref{Fsedvisidens17}.
The various flux contributions at 2.11\,$\mu$m  
are 62.2\% stellar light, 26.1\% scattered radiation and 
10.7\% dust emission (see Fig.~\ref{Fffraction17}), i.e. the radiation
emitted by the circumstellar shell itself consists of 71\% scattered radiation
and 29\% direct dust emission.
The  shell's model intensity distribution is shown in
Fig.~\ref{Fffraction17} and was found to be ring-like.
This appears to be typical for optically thin shells
(here $\tau_{0.55\mu{\rm m}}=7$, $\tau_{2.11\mu{\rm m}}=0.55$)
showing limb-brightened dust-condensation zones.
Accordingly, the interpretation of the observational data by FWHM Gau\ss\
diameters may give misleading results.
The ring diameter is equal to the inner diameter of the hot shell
($\sim 69$\,mas), and the diameter of the central star amounts to
$\sim 1$\,mas.
The bolometric flux, $F_{\rm bol}$, is
$8.17 \cdot 10^{-10}$\,Wm$^{-2}$ corresponding to a central-star 
luminosity of $L/L_{\odot} = 25\,462 \cdot (d/{\rm kpc})^{2}$.

This two-component model  can be interpreted in terms of a termination of an
enhanced mass-loss phase roughly 90 yr (for $d=5$\,kpc) ago.
The assumption that IRC\,+10\,420 had passed through a
superwind phase in its recent history is in line with
its evolutionary status of
an object in transition from the Red-Supergiant to the Wolf-Rayet phase.
The mass-loss rates of
the components can be determined to be
$\dot{M}_{1}= 7.0\ 10^{-5}$\,$M_{\odot}/{\rm yr}$ and
$\dot{M}_{2}= 1.1\ 10^{-3}$\,$M_{\odot}/{\rm yr}$.
\section{Interferometry with the VLTI and the AMBER instrument}
\begin{figure}
\begin{center}
  \begin{tabular}{c}
    \psfig{figure=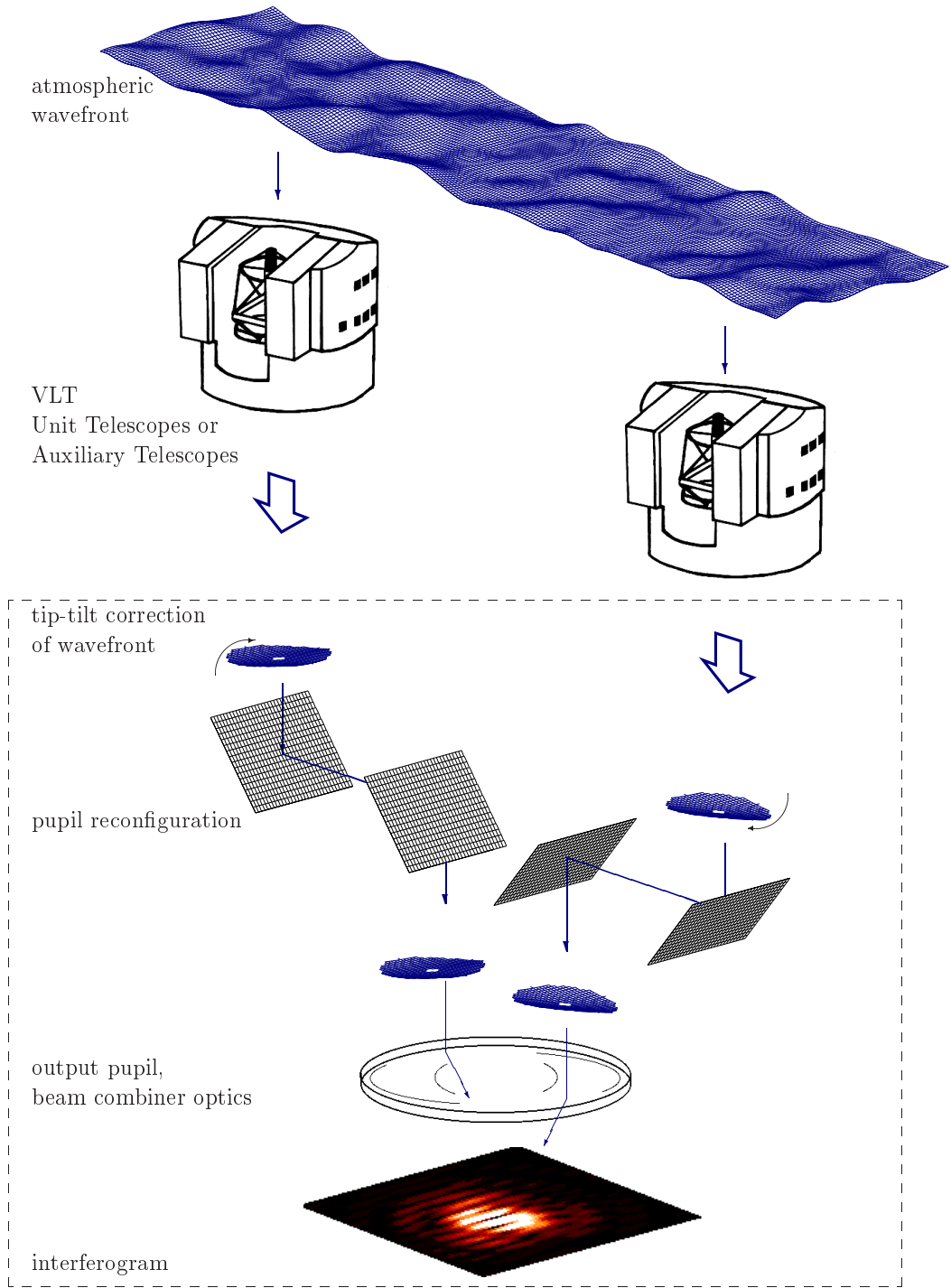,height=0.9\textheight}
\end{tabular}
\end{center}
\caption[fluss3d]
{
Schematic view of interferometric imaging with the VLTI AMBER
 camera.
} \label{Ffluss3d}
\end{figure}

In the previous section, the spatial intensity profile of the dusty supergiant
IRC\,+10\,420 (see Fig.~\ref{Fffraction17}) was derived by means of radiative
transfer models and their comparison with photometric and interferometric
observations. This $2.11\,\mu$m intensity profile will serve as object
intensity profile in the simulation of monochromatic VLTI observations. 
The next steps are the simulation of  light propagation from the object to
the detector (through  atmosphere, telescopes, and the AMBER wide-field mode
instrument),
simulation of photon noise and detector read-out noise, and finally data
processing of the interferograms.
A schematic view of Michelson interferometry with the VLTI and the AMBER camera
is given in Fig.~\ref{Ffluss3d}.
\subsection{Computer simulation of interferometric imaging}
\begin{figure}
\begin{center}
  \begin{tabular}{c}
    \psfig{figure=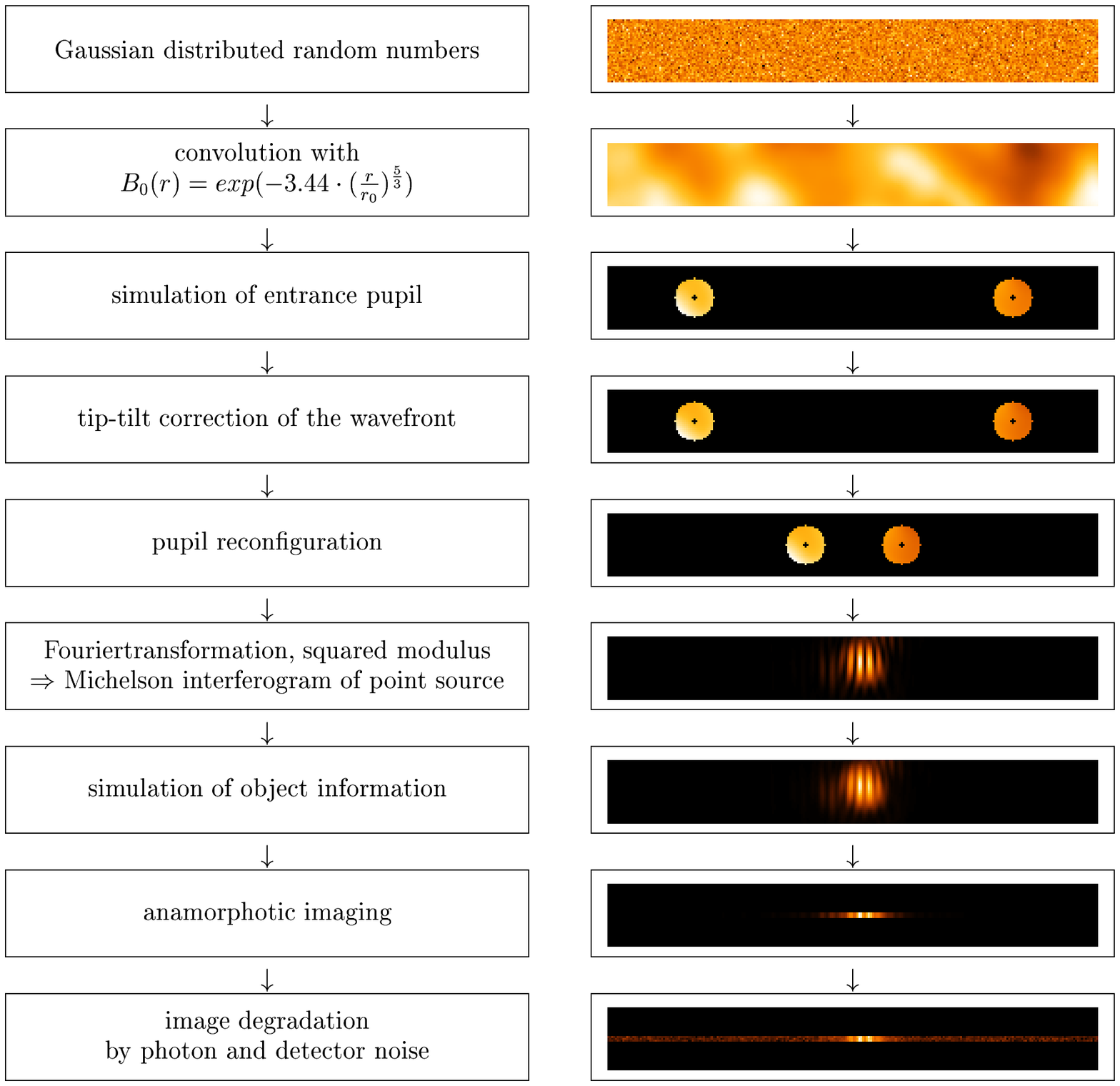,width=1.0\textwidth}
\end{tabular}
\end{center}
\caption[fluss2d]
{
Flow chart of the simulation of VLTI AMBER interferograms
(wide-field mode, i.e. without fiber optics spatial filtering).
} \label{Ffluss2d}
\end{figure}
Fig.~\ref{Ffluss2d} shows a flow chart of our simulation of interferometric
imaging with the VLTI
(ATs, or UTs with adaptive optics)
and the AMBER camera in the wide-field
mode (i.e. without fiber optics spatial filtering).
In a first step, an array of
Gaussian distributed random numbers is generated and convolved  with the
correlation function of the atmospheric refraction index variations
in order to generate wavefronts degraded by atmospheric
turbulence\cite{Rod81}.
The typical size of the atmospheric turbulence cells is given
by the Fried parameter $r_{0}$. After the simulation of the entrance pupil,
the next step incorporates
the tip-tilt correction
%%In the simulations, image motion of the dominant speckle is corrected,
but we allow for a residual tip-tilt  error $\delta_{\rm tt}$.
In the next step a typical Michelson output pupil is simulated (pupil
reconfiguration).
The output pupil is chosen such that (i) in the optical transfer function
the off-axis peaks are separated from the central peak, and (ii)
the interferograms are sampled with the smallest number of pixels to assure
the lowest influence of detector noise.
The following step includes
light propagation through the beam combiner lens
to the focal plane. The squared modulus of  the
Fourier transform of the complex amplitude in front of the beam combiner lens
yields the intensity distribution of a Michelson interferogram of a point
source.
In the next step the required object intensity distribution is simulated
(given here by the intensity distribution of IRC\,+10\,420)
to obtain the Michelson interferogram of the object:
The Fourier transformation of the object intensity distribution, calculated
at those spatial frequencies covered by the the simulated interferometer
baseline vector, is multiplied with the off-axis peaks of the transfer
function of the generated Michelson interferogram\cite{Tall2}.
Finally, Poisson photon-noise and detector read-out noise is injected to the
interferograms.
The noise level depends, among other parameters, on the
number of detectable photons,
%%%given here by the $K$ magnitude of the IRC\,+10\,420,
the total optical throughput of the
interferometer and the quantum efficiency of the detector.
Details\cite{MalEtal2000}
are shown in Table~\ref{Tsimu1}.
\begin{table} %%% [h]  
\begin{center}       
\caption{Parameters of simulations} \vspace*{1ex}
\label{Tsimu1}
\begin{tabular}{|c|c|c|c|c|c|c|}
%%% use of \rule[]{}{} below opens up each row
\hline
\rule[-1ex]{0pt}{3.5ex} wavelength   & brightness    & opt.\ throughput &
    read-out noise & quantum efficiency & exposure & photon number  
\\ \hline 
\rule[-1ex]{0pt}{3.5ex} $2.11\,\mu$m & $K=3.5$\,mag  & 0.108          &
    15\,e$^{-}$    &  0.6               & 50\,ms   & $ 4.42 \cdot 10^{6}$
\\ \hline
\end{tabular}
\end{center}
\end{table}

\subsection{Visibility data for IRC\,+10\,420}
\begin{figure}
  \begin{tabular}{c}
    \psfig{figure=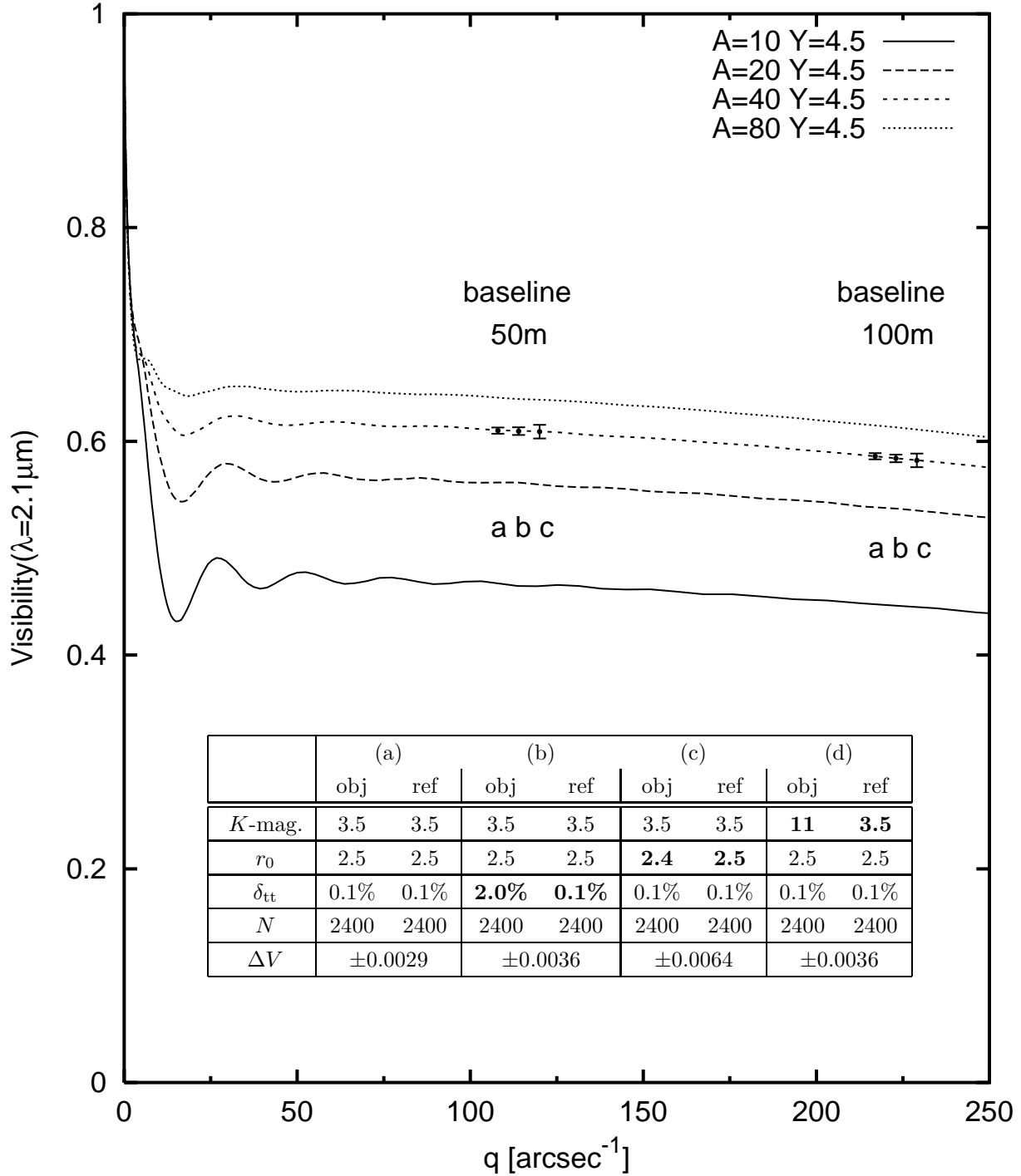,width=0.94\textwidth} \\
  \end{tabular}
\caption[simvisATH]
{
Dependence of the error bars of simulated AT-VLTI/AMBER 
(wide-field mode) observations of IRC\,+10\,420 at 2.11\,$\mu$m on
(i) seeing differences of object and reference star
     observations (Fried parameter, $r_{0}$, differences);
(ii) different residual tip-tilt errors, $\delta_{\rm tt}$,
of object and reference star observations; and
(iii) object brightness. 
Lines refer to radiative transfer models of
different superwind amplitudes $A$. The symbols refer to 
AT-VLTI/AMBER simulations (each with $N$=2400 interferograms)
of the $A$=40 intensity profile
(see Fig.~\ref{Fffraction17}) for baselines of 50 and 100\,m
%%%%%(100\,m $\stackrel{\wedge}{=}$ $q$=227 arc\,sec$^{-1}$).
(100\,m: $q$=227 arc\,sec$^{-1}$).
To better distinguish between the simulations,
the data points belonging to one baseline are somewhat shifted 
with respect to the spatial frequency. % (in units of 6 arc\,sec$^{-1}$).
The error bars are based on a triple iteration of each simulation.
The table gives the parameters of the simulations (a) to (d). Note
that though simulation\,(d) corresponds to an IRC\,+10\,420-like intensity
distribution, it refers to a much fainter object of $K$=11\,mag (instead of
3.5\,mag).
} \label{FsimvisATH}
%%%%%%%%%%%%%%%%%%%%%% Insert TeX Table into the Figure %%%%%%%%%%%%%%%%%%
\hspace*{0.3cm}
\begin{minipage}{1.0\textwidth}
\vspace*{-19cm}
%\begin{table}[h]  
\begin{center}      
\begin{tabular}{|c|cc|cc|cc|cc|}
%%% use of \rule[]{}{} below opens up each row
\hline
\rule[-1ex]{0pt}{3.5ex}  &
                        \multicolumn{2}{c}{(a)} &
                        \multicolumn{2}{c}{(b)} & 
                        \multicolumn{2}{c}{(c)} & 
                        \multicolumn{2}{c|}{(d)} 
\\  
\rule[-1ex]{0pt}{3.5ex}  &
                         obj & ref & 
                         obj & ref & 
                         obj & ref & 
                         obj & ref  
\\ \hline \hline
\rule[-1ex]{0pt}{3.5ex} $K$-mag. &
                         3.5  &  3.5  &
                         3.5  &  3.5  &
                         3.5  &  3.5  &
                   {\bf 11}   &  {\bf 3.5}  
\\ \hline
\rule[-1ex]{0pt}{3.5ex} $r_{0}$ &
                         2.5  &  2.5  &
                         2.5  &  2.5  &
                   {\bf 2.4}  &  {\bf 2.5}  &
                         2.5  &  2.5  
\\ \hline
\rule[-1ex]{0pt}{3.5ex} $\delta_{\rm tt}$ &
                       0.1\%   & 0.1\%   & 
                   {\bf 2.0\%} & {\bf 0.1\%}   & 
                       0.1\%   & 0.1\%   & 
                       0.1\%   & 0.1\%    
\\ \hline
\rule[-1ex]{0pt}{3.5ex} $N$ &
                        2400   & 2400   &
                        2400   & 2400   &
                        2400   & 2400   &
                        2400   & 2400   
\\ \hline
\rule[-1ex]{0pt}{3.5ex} $\Delta V$ &
                        \multicolumn{2}{c|}{$\pm 0.0029$} &
                        \multicolumn{2}{c|}{$\pm 0.0036$} &
                        \multicolumn{2}{c|}{$\pm 0.0064$} &
                        \multicolumn{2}{c|}{$\pm 0.0036$} 
\\ \hline
\end{tabular}
\end{center}
%\end{table}
\end{minipage}
\end{figure}

We performed simulations of IRC\,+10\,420 AT visibility observations
in the AMBER wide-field mode
and studied the influence of various observational parameters on the
visibility accuracy.
Visibility error bars were, for example, obtained for the following
observational parameters: different seeing during the observation of
object and reference star (Fried parameters $r_{0,{\rm object}}$=2.4\,m,
$r_{0,{\rm ref.}}$=2.5\,m), different residual tip-tilt error
($\delta_{\rm tt,object}$=2\% of the Airy disk diameter,
$\delta_{\rm tt,ref.}$=0.1\%), and object 
brightness ($K_{\rm object}$=3.5\,mag and 11\,mag, $K_{\rm ref.}$=3.5\,mag).
In the computer experiments (a)-(c),
object and reference star were assumed to have
the same brightness ($K$=3.5\,mag, see Table~\ref{Tsimu1}),
in experiment (d) fainter objects  ($K$=11\,mag)
were simulated as well.
Fig.~\ref{FsimvisATH} shows the results of the simulations (a) to (c)
based on the $A=40$ intensity profile (see Fig.~\ref{Fffraction17})
for baselines of 50 and 100\,m together with the model predictions for
IRC\,+10\,420. To obtain error bars each simulation was repeated three times.
%%The error bar is given by twice the value of the largest deviation from
%%the model visibility ($A=40$).
The insetted table lists the parameters of the simulations.

Simulation (a) based on 2400 interferograms
represents the ideal case of excellent seeing conditions
($r_{0,{\rm object}}$=2.5\,m, $r_{0,{\rm ref.}}$=2.5\,m)
and an almost perfect tip-tilt correction (residual tip-tilt error
$\delta_{\rm tt}=0.1$\%).
The visibility error $\Delta V$ amounts to $\pm 0.0029$.
Simulation (b) illustrates the influence of a larger residual object 
tip-tilt error ($\delta_{\rm tt}$=2\%) leading to
$\Delta V \pm 0.0036$. 
Simulation (c) shows the impact of different seeing conditions for
object and reference star. 
%%(($f_{\rm S}=0.62$) increasing the visibiliy error bar by a factor of 2
%%($\Delta V \pm 0.0064$). The influence of the number of expsoures
%%is demonstrated by simulation (d) using only half of the interferograms
%%($N=1200$) which gives $\Delta V \pm 0.0068$. 
%
In the last simulation (d) of this series (shown only in the insetted table of
Fig.~\ref{FsimvisATH})
an IRC\,+10\,420-like intensity distribution is assumed but the simulated
$K$-magnitude is 11\,mag  (instead of 3.5\,mag).
Although the photon number decreases to $N=4.41 \cdot 10^{3}$
(i.e.\ by a factor of 1000, see  Table~\ref{Tsimu1}), the visibility
accuracy is still high, i.e.\ $\Delta V \pm 0.0036$. 
%
%
%
%%%%%%%%%%%%%%%%%%%%%%%%%%%%%%%%%%%%%%%%%%%%%%%%%%%%%%%%%%%%%%%%%%%%%%%%%
%%%%%%%%%%%%%%%%%%%%%%%%%%%%%
%
\section{Conclusions}
We have presented computer simulations of interferometric imaging with the
VLT interferometer and the AMBER instrument in the wide-field
mode. These simulations include both the
astrophysical modelling of a stellar object by radiative transfer
calculations and the simulation of  light propagation from the object to the
detector and simulation of photon noise and detector read-out noise.
We focussed on stars in late stages of stellar evolution and examplarily
studied one of its most outstanding representatives, the dusty supergiant
IRC\,+10\,420. The model intensity distribution of this key object,
obtained by radiative transfer calculations,
%%%constrained by photometric and speckle interferometric observations,
served as astrophyiscal input for the VLTI/AMBER simulations.

The results of these simulations show the dependence of the
visibility error bar on various observational parameters.
With these simulations at hand one can immediately see under
which conditions the visibility data quality would allow us to discriminate
between different model assumptions
(e.g.\ the size of the superwind amplitude).
%%This appears crucial for the
%%verification and improvement of existing models and thus for our
%%%understanding of supergiants in general.
Inspection of
Fig.~\ref{FsimvisATH} shows that in all studied cases 
the observations will give clear preference to one particular model.
Therefore, observations with VLTI will certainly be well suited to gain
deeper insight into the physics of dusty supergiants.

\acknowledgments     %>>>> equivalent to \section*{ACKNOWLEDGMENTS}       

The bispectrum speckle observations were made with
the SAO 6\,m telescope operated by the
Special Astrophysical Observatory, Russia.
The radiative-transfer calculations are based on 
the code DUSTY developed by \v{Z}.\ Ivezi\'c, M.\ Nenkova and M. Elitzur.

%xxxxx


\begin{thebibliography}{99}   
%% the last item specifies width of reference number column

\bibitem{GlinEtal2000}
 Glindemann, A., et al., 2000, Proc. SPIE Conf. 4006-01

\bibitem{PetEtal98}
 Petrov, R., Malbet, F., Richichi, A., Hofmann, K.-H.: 1998,
   ESO Messenger 92, 11

\bibitem{PetEtal2000}
 Petrov, R., et al. , 2000, Proc. SPIE Conf. 4006-07

\bibitem{RichEtal2000}
 Richichi, A., et al., 2000, Proc. SPIE Conf. 4006-08

\bibitem{HumStrMurLow73} 
    Humphreys R.M., Strecker D.W., Murdock T.L., Low, F.J., 1973, ApJ 179, L49

\bibitem{OudGroeMatBloSah96}
    Oudmaijer R.D., Groenewegen M.A.T., Matthews H.E., Blommaert J.A.D.L,
    Sahu K.C., 1996, MNRAS 280, 1062

\bibitem{KloChePan97}
    Klochkova\,V.G., Chentsov\,E.L., Panchuk\,,V.E., 1997, MNRAS\,292,19

\bibitem{GigWooWeb76}
    Giguere P.T., Woolf N.J., Webber J.C., 1976, ApJ 207, L195

\bibitem{MutEtal79}
    Mutel R.L., Fix J.D., Benson J.M., Webber J.C., 1979, ApJ 228, 771

\bibitem{NedBow92}
    Nedoluha G.E., Bowers P.F., 1992, ApJ 392, 249

\bibitem{KnaMor85}
    Knapp G.R., Morris M., 1985, ApJ 292, 640

\bibitem{JonHumGehEtal93}
    Jones T.J., Humphreys R.M, Gehrz, R.D. et al., 1993, ApJ 411, 323

\bibitem{HumSmiDavEtal97} 
    Humphreys R.M., Smith N., Davidson K. et al., 1997, AJ 114, 2778

\bibitem{OudEtal94}
    Oudmaijer R.D., Geballe T.R., Waters, L.B.F.M, Sahu K.C., 1994,
    A\&A 281, L33

\bibitem{Oud98}
     Oudmaijer R.D., 1998, A\&AS 129, 541

\bibitem{DyckEtal84}
 Dyck\,H., Zuckerman\,B., Leinert\,C., Beckwith\,S.,\,1984,\,ApJ\,287,\,801

\bibitem{RidgEtal86}
     Ridgway S.T., Joyce R.R., Connors D., Pipher J.L., Dainty C., 1986,
     ApJ 302, 662

\bibitem{CobFix87}
    Cobb M.L., Fix J.D., 1987, ApJ 315, 325

\bibitem{ChrEtal90}
Christou J.C., Ridgway S.T., Buscher D.F., Haniff C.A., McCarthy~Jr. D.W.,
  1990, Astrophysics with infrared arrays, R.\,Elston (ed.),
  ASP conf.\ series 14, p.~133

\bibitem{KastWein95}
     Kastner J., Weintraub D.A., 1995, ApJ 452, 833

\bibitem{BloeckEtal99}
 Bl\"ocker T., Balega Y., Hofmann K.-H., Lichtenth\"aler J., Osterbart R.,
 Weigelt G., 1999, A\&A 348, 805

\bibitem{Wei77}
     Weigelt G., 1977, Optics Commun. 21, 55

\bibitem{LohWeiWir83}
    Lohmann A.W., Weigelt G., Wirnitzer B., 1983, Appl. Opt. 22, 4028

\bibitem{HofWei86}
     Hofmann K.-H., Weigelt G., 1986, A\&A 167, L15

\bibitem{Lab70}
     Labeyrie A., 1970, A\&A  6, 85

\bibitem{WalEtal91}
    Walmsley C.M., Chini R., Kreysa E. et al.,
    1991, A\&A 248, 555

\bibitem{CraEtal76}
    Craine E.R., Schuster W.J., Tapia S., Vrba F.J., 1976, ApJ 205, 802

\bibitem{SavMat79}
    Savage B.D., Mathis J.S., 1979, ARA\&A 17, 73

\bibitem{IveNenEli97}
    Ivezi\'c \v{Z}., Nenkova M. Elitzur M., 1997, User Manual for DUSTY,
    University of Kentucky %%%, accessible at
    (http://www.pa.uky.edu/\~moshe/dusty)

\bibitem{IveEli97}
    Ivezi\'c \v{Z}., Elitzur M., 1997, MNRAS 287, 799

\bibitem{MRN77}
    Mathis J.S., Rumpl W., Nordsieck K.H., 1977, ApJ 217, 425 
    
\bibitem{DraLee84}
    Draine B.T., Lee H.M., 1984, ApJ 285, 89

\bibitem{Rod81}
    Roddier F., 1981, Progress in Optics XIX, 281

\bibitem{Tall2}
   Tallon M., Tallon-Bosc I., 1992, A\&A 253, 641

\bibitem{MalEtal2000}
   Malbet F., et al., 2000, VLTI AMBER Instrument Analysis Review.

\end{thebibliography}
\end{document}